\documentclass[epj,nopacs]{svjour}
\usepackage{latexsym}
\usepackage{graphics}
\begin{document}
\title{From Regge Behavior to DGLAP Evolution}
\author{L. Csernai\inst{1} \and L. Jenkovszky\inst{2}
\and K. Kontros\inst{3} \and A. Lengyel\inst{3} \and V.
Magas\inst{4} \and F. Paccanoni\inst{5}}

\institute{Section for Theoretical Physics, Department of
Physics, University of Bergen, Allegaten 55, 5007 Bergen, Norway
(\email{csernai@fi.uib.no}) \and Bogolyubov Institute for
Theoretical Physics, National Academy of Sciences of Ukraine,
01143 Kiev, Ukraine (\email{jenk@gluk.org}) \and Institute of
Electron Physics, National Academy of Sciences of Ukraine,
Universitetska 21, 88016 Uzhgorod, Ukraine
(\email{jeno@kontr.uzhgorod.ua, sasha@len.uzhgorod.ua}) \and
Center for Physics of Fundamental Interactions, Instituto Superior
Tecnico, Av. Rovisco Pais, 1049-001 Lisbon, Portugal
(\email{vladimir@cfif.ist.utl.pt}) \and Dipartimento di Fisica,
Universit\`a di Padova, Instituto Nazionale di Fisica Nucleare,
Sezione di Padova via F. Marzolo 8, I-35131 Padova, Italy
(\email{paccanoni@pd.infn.it})}

\date{Received: date / Revised version: date}

\abstract{We study the interface between Regge behavior and DGLAP
evolution in a non-perturbative model for the nucleon structure
function based on a multipole pomeron exchange. This model
provides the input for a subsequent DGLAP evolution that we
calculate numerically. The soft input and its evolution give a
good fit to the experimental data in the whole available range of
$x$ and $Q^2$.
 }

\maketitle

\section{Introduction}
\label{intro}

There is a growing consensus on that nucleon structure functions
(SF) at small virtualities $Q^2$ are Regge behaved and that at
large $Q^2$ they follow QCD evolution \cite{BGJPP}. The two
regimes are incompatible because the DGLAP evolution \cite{DGLAP},
in general, changes the functional form of any Regge behavior
(Regge models compatible with QCD evolution in a limited range of
$Q^2$ will be mentioned below). What remains completely unclear
is the border between the two.

Phenomenologically, the Regge pole approach to deep inelastic
scattering implies that the structure functions are sums of
powers in $x$, modulus logarithmic terms, each with a
$Q^2-$dependent residue factor. The rapid rise in $Q^2$ of the
structure functions, observed at HERA, initially was considered
as a sign of departure from the standard Regge behavior. The
reason was that the data, when fitted by a single ``Regge-pomeron"
term $\sim x^{-\lambda}$, where $\lambda$ is the pomeron
intercept minus one, show that $\lambda={d\over{d\ln 1/x}}\ln
F_2(x,Q^2)$ definitely rises with $Q^2$. If so, Regge
factorisation should be broken, since the ``effective" pomeron
trajectory appears to be $Q^2-$dependent.

The above categoric conclusion was however based on a
parametrisation of the SF by a single pomeron term, e.g. $\sim
x^{-\lambda}$. It is evident, that additional, decreasing
(``subleading") terms, typical of any Regge pole model, will
modify the properties of the leading one (pomeron). Moreover, by
increasing the number of the free parameters and with complicated
residue functions one may fit the data for arbitrarily large
values of $Q^2$ in a factorised form, without introducing any
$Q^2-$dependent trajectory, although it makes little sense to
reach good $\chi^2$ at any rate with an indefinite number of free
parameters introduced without any physical justification.

The (in)dependence of the Regge behavior (trajectory) from the
external masses (virtualities) comes from the following
arguments. The Regge asymptotic behavior
\begin{equation}
\label{eq2} A(s,t,Q^2)=\xi(t)\beta(t,Q^2)\Bigl({s\over
{s_0}}\Bigr)^ {\alpha(t)},
\end{equation}
where $\xi(t)$ is the signature factor, $\beta(t,Q^2)$ is the
residue function, $\alpha(t)$ is the Regge trajectory and $s,t$
are the Mandelstam variables, comes from a Sommerfeld-Watson
transform over the partial-wave amplitude (see for example
\cite{Collins})

\begin{equation}
\label{eq3} a(l,t,Q^2)={\beta(t,Q^2)\over{l-\alpha(t)}}.
\end{equation}

By Regge factorisation, $Q^2-$dependence can be introduced in the
residue, but not in the trajectory.

In this paper we are mainly interested in the small-$x$ or
high-energy behavior of the structure functions and photon-proton
total cross sections, therefore we concentrate on the pomeron (in
terms of the Regge pole model) or singlet component of the SF,
although, clearly, good fits and the evolution equation require
the whole kinematical region in $x$ to be properly covered. Model
for the pomeron, or singlet SF, and non leading contributions, or
non-singlet SF, will be presented in Sec. \ref{sec:2}.

Even though the role of the QCD evolution was studied in a great
number of papers (see e.g. \cite{CDL} and references therein) the
border and the interface between the non-perturbative Regge
dynamics and perturbative QCD evolution is still determined
merely by trial and error. Depending on the scope of the study,
one may prefer either approximate solutions
\cite{Paccanoni,JKP,JLP} of the DGLAP equation \cite{DGLAP} or
computer calculations, for which purpose efficient numerical
solutions and relevant computer codes are available \cite{MK}.

Explicit analytical solutions are attractive for their simplicity
and physical transparency. In paper \cite{JKP}, for example, the
transition due to the evolution from a logarithmic rise in $x$
(dipole model) to a power-like was shown explicitly by means of
an approximate solution of the DGLAP equation. Another example,
in the Regge framework, is given by the CKMT model
\cite{CKMT,KMP}.

Of particular interest are self-consistent solutions of the DGLAP
evolution equations \cite{Paccanoni}, i.e. those whose functional
form does not change under the evolution. The product of
logarithms (in $x$ and in $Q^2$) proved to be particularly
stable. In ref. \cite{JLP} this stability for the calculated
$Q^2-$dependent coefficients, appearing in front of the factors
logarithmic in $x$, was checked experimentally. The stability of
the combination of the logarithms
\begin{equation}
\label{eq4} F_2(x,Q^2)=a+m\xi, \quad
\xi=\ln{Q^2\over{Q_0^2}}\ln{x_0\over{x}}
\end{equation}
was noticed also by Buchm\"uller and Haidt \cite{Haidt} from
phenomenological analyses of the data. Ref. \cite{CDL} approaches
this problem in a different spirit. The inclusion of a
"hard-pomeron" term, with Regge intercept about $1.4$, allows for
a fit to structure functions with a $\chi^2$ per data point near
$1$. Moreover this new term is self-consistent, in the sense that
a phenomenological parametrization of its $Q^2$-dependence, at
small $Q^2$, turns out to be in agreement with perturbative
evolution. However, the resulting dynamics here is different from
the one necessary to describe hadron-hadron scattering.

Aiming at an unbiased extraction of the $Q^2-$dependent factors
appearing in the structure functions, the authors in \cite{DJLP}
have fitted them for a number of different models of the pomeron,
each appended by subleading terms as well as a large-$x$ factor.
The numerical values of the fitted coefficients for a large
number of fixed values of $Q^2$ (Figs. 1, 2, 3 in \cite{DJLP}, see
also Fig. 1 in ref. \cite{CS}) can be used as ``experimental
values", independent of any evolution scheme.

The aim of this paper is to check the onset of QCD evolution for
the model of small-$x$ SF that was discussed earlier in
\cite{JLP,DJLP}. This model contains a small (minimal in our
opinion) number of free parameters, still it is feasible and,
supplemented by an evolution scheme, may have numerous
applications.

Our approach concerns the class of phenomenological
parametrisation of structure functions describing the
experimental data in the non-perturbative region; using this
parametrisation as input to the DGLAP equations one obtains a
description of the experiment in the whole kinematic range
\cite{CKMT,CDL}. This genuine evolution, in contrast to global
fits \cite{ACS,MRST}, does not use data at large $Q^{2}$ to
constrain the fit already at small $Q^{2}$. In comparison to the
DL \cite{CDL} model, where the pomeron has a hard nature, we
choose a multipole pomeron with unit intercept.

In the present paper the evolution is calculated numerically by
means of the codes developed and published by Miyama and Kumano
\cite{MK}.

\section{Multipole pomeron model}
\label{sec:2}

We analyze a unit intercept Multipole pomeron, rather than a
supercritical one as in the CKMT case \cite{CKMT}. Here Multipole
pomeron (MP) means that the pomeron is a multipole instead of
just a simple pole. The advantages of this approach, both for
hadron-hadron and for lepton-hadron scattering, were tested and
discussed in numerous papers on that subject (see
\cite{BGJPP,JLP,DJLP,CS,JPP,DGJP,DJP,DLM2} and references
therein). The main point is that rising cross sections (and
rising SF) can be produced with a unit pomeron intercept. The
number and the relative weight of the contributing multipoles is
a very important question. In QCD it was studied in ref.
\cite{FJKLPP}. The conclusion of both the theory \cite{FJKLPP}
and phenomenology is that a limited (moreover, small!) number of
multipoles is sufficient in the present energy (or $x, Q^2$)
range. In fact, a dipole (logarithmic rise), eventually with a
minor tripole contribution (squared logarithm) accounts for the
large part of the data (higher order poles do not manifest).

The dipole (and tripole) pomeron are typically ``soft" objects.
For the total cross sections they give fits close to those of a
``soft" supercritical, i.e. with $\alpha(0)\approx 1.08$,
pomeron. However, with an extra variable -- $t$ in hadron-hadron
processes or $Q^2$ in lepton-hadron reactions -- the differences
become essential. The result of the analysis below, including QCD
evolution, makes this comparison more complete.

Both non-perturbative inputs are written as sums of singlet and
non-singlet terms:

\begin{equation}
\label{eq5} F_{2}(x,Q^{2})=F_{S}(x,Q^{2})+F_{NS}(x,Q^{2}).
\end{equation}
We use the same non-singlet term taking it from ref. \cite{KMP} as

\begin{equation}
\label{eq6} F_{NS}(x,Q^{2})=D\cdot x^{1-\alpha_{R}}\cdot
(1-x)^{n(Q^{2})}\cdot
\left(\frac{Q^{2}}{Q^{2}+b}\right)^{\alpha_{R}},
\end{equation}

\begin{equation}
\label{eq7} n(Q^{2})=\frac{3}{2} \cdot
\left(1+\frac{Q^{2}}{Q^{2}+c}\right),
\end{equation}
where, contrary to the original paper \cite{KMP}, we disregard
the difference between the light quarks.

The singlet component of the SF, corresponding to a multipole
(single+double+triple) pomeron is a sum of logarithms:

\begin{equation}
\begin{array}{cc}
\label{eq8} F_{S}(x,Q^{2})=Q^{2}
\left[A\left(\frac{a}{a+Q^{2}}\right)^{\alpha}+B\left(\frac{a}
{a+Q^{2}}\right)^{\beta} \log{\left(\frac{Q^{2}}{x}\right)}+ \right. \\
\left. C\left(\frac{a}{a+Q^{2}}\right)^{\gamma}
\log^2{\left(\frac{Q^{2}}{x}\right)}\right] (1-x)^{n(Q^{2})+4}.
\end{array}
\end{equation}
The same type of singlet component was used in \cite{CS}.

The real photon-proton total cross-section has the following form:
\begin{equation}
\begin{array}{cc}
\label{eq9} \sigma^{tot}_{\gamma p}(W^{2})=4 \pi^{2} \alpha_{EM}
\cdot \left( A+B \cdot \log(W^{2}_{1})+ \right.
\\ \left. C \cdot \log^2(W^{2}_{1}) +D \cdot
d^{-\alpha_{R}} \cdot (W^{2}_{1})^{\alpha_{R}-1} \right),
\end{array}
\end{equation}
where $W^{2}_{1}=W^{2}-m_{p}^{2}$.

The parameters of (\ref{eq5})-(\ref{eq9}) were determined from a
combined fit of $\sigma^{tot}_{\gamma p}$ for $W>4$ $GeV$ and
structure function in the range of $0.045GeV^{2}\leq Q^{2}\leq
5GeV^{2}$.

We performed these fits to the set of the experimental data
\cite{data} using eqs. (\ref{eq5})-(\ref{eq9}). The values of the
fitted parameters and of $\chi^2$ are quoted in Table 1.

\begin{table} \caption{Parameters of the MP model} \label{tab:1}
\begin{tabular}{cc}
\hline\noalign{\smallskip}
MP model & values of the parameters \\
\noalign{\smallskip}\hline\noalign{\smallskip}
$A$ & $0.8286$ \\
$B$ & $-0.1602.10^{-1}$ \\
$C$ & $0.8208.10^{-2}$ \\
$a$ & $0.7951$ \\
$b$ & $1.263$ \\
$c$ & $4.430$ \\
$\alpha$ & $2.0$ (fixed) \\
$\beta$ & $2.0$ (fixed)\\
$\gamma$ & $0.7711$ \\
$\alpha_{R}$ & $0.415$ (fixed) \\
$D$ & $1.434$ \\
$\chi^{2}/dof$ & $1.08$ \\
\noalign{\smallskip}\hline
\end{tabular}
\end{table}

We have a rather small (eight) number of free parameters. In the
MP model we fix the reggeon intercept as in the CKMT model.
Additionally $\alpha$ and $\beta$ were fixed being not sensitive
to the fit. The resulting fits are shown in Figs. 1 and 2.

As already mentioned, we chose the starting point for the QCD
evolution in such a way as to have best fits with the smallest
number of the free parameters. By trial we found that the value
$Q_{0}^{2}=6\;GeV^{2}$ is preferable. In the numerical
calculation, we assume that the only difference between the
sea-quark and gluon distributions is a factor $G(1-x)^{-2}$ in
accordance with the dimensional counting rules. The value of the
constant $G$ is $3.8$, the number of flavors is $4$,
$\lambda=200\;MeV$. In contrast to \cite{CDL}, in our case the
ratio of the gluon distribution to the singlet quark distribution
is smaller by a factor of two.

Figs. 2 and 3 show the extrapolation of the $Q^2-$depen-dent
``soft" Regge input as well as the results of our numerical
calculations of the DGLAP equation by means of the NLO
brute-force method, developed by Miyama and Kumano in ref.
\cite{MK}, thus extended to the high values of $Q^2$, measured at
\cite{data}. The input coincides with the structure function
itself, supplemented by the aforesaid gluon distribution, since
the corrections implied by the $\overline{MS}$ scheme are small
and we neglected them. Although we are primarily interested in
the small-$x$ behavior, large-$x$ data, including those from
fixed-target experiments \cite{data} were also taken into account
since they can influence the small-$x$ behavior through
integration in the evolution equation.

\begin{figure}
\resizebox{0.5\textwidth}{!}{\includegraphics{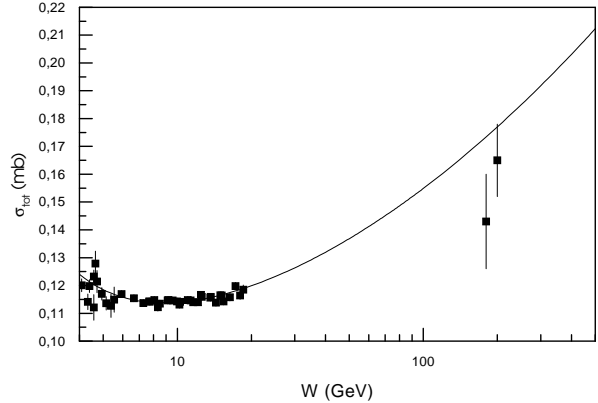}}
\caption{Total $\gamma p$ cross-section versus $W$ in the MP
model.} \label{fig:1}
\end{figure}

\section{Conclusions}
\label{sec:3}

We show that the world's data on $F_2$, down to $Q^2=0$, are
reasonably well described within the Regge model we have
considered. The model succeed in reproducing the data for all $x$
and a large $Q^2$ interval, until $Q^2\approx 6\;GeV^2$,
whereupon DGLAP equations must be used in order to describe
correctly the $Q^2$ evolution. What happens at the border between
Regge regime and QCD evolution is clarified in the last icon of
Fig. 2 and the first three frames of Fig. 3. While the starting
point for the DGLAP evolution has been chosen at
$Q_0^2=6\;GeV^2$, the extrapolation of the ``soft" Regge behavior
gives a satisfactory fit of the data until $Q^2\approx
10\;GeV^2$. In Fig.3, one can see that the fit, based on eqs.
(\ref{eq5})-(\ref{eq9}), deteriorates when $Q^2$ increases and, at
$Q^2=20\;GeV^2$, it is excluded by the data. There is however a
region, a $Q^2$ interval, where the two regimes, Regge and
perturbative QCD, are compatible. The comparison with other
models, that consider only the small-$x$ region
\cite{CDL,Paccanoni,JKP,JLP,Haidt}, shows that the region of
compatibility shrinks when all the $x$-range is taken into
account. To what extent this depends on the specific model we
have chosen, is an interesting question and will be studied
elsewhere.

To summarize, we find that the model described in eq.
(\ref{eq8}), for singlet SF, supplemented by a non-singlet
component as in eq. (\ref{eq6}), together with the DGLAP
evolution, provides a simple and economic solution that could be
useful for further practical applications, for example in nuclear
physics. In addition, we can conclude that the soft input and its
evolution give a good fit to the experimental data in whole
available range of the variables $x$ and $Q^2$.

\begin{acknowledgement}
K. Kontros and A. Lengyel acknowledge the hospitality and support
at the \texttt{Bergen Computational Physics Laboratory} (BCPL),
where this work was started, while L. Jenkovszky thanks
University and the INFN section in Padova, where it was completed.
\end{acknowledgement}

\begin{figure*}
\resizebox{0.9\textwidth}{!}{\includegraphics{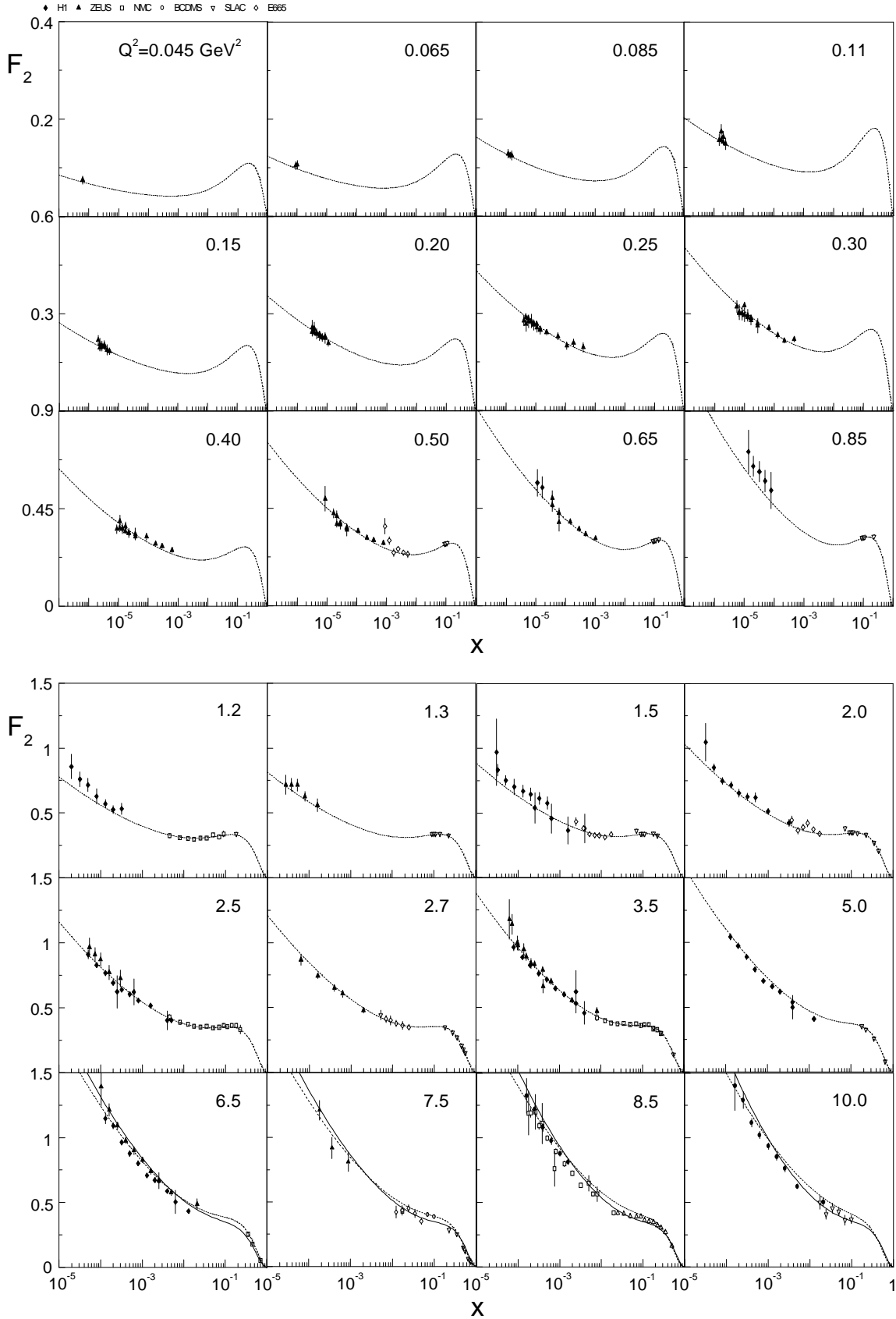}}
\caption{Fit of $F_{2}$ for the smallest available values of
$Q^{2}$ up to $5$ $GeV^{2}$ and its extrapolation (dashed curve).
The solid curve displays the result of evolution.} \label{fig:2}
\end{figure*}

\begin{figure*}
\resizebox{0.9\textwidth}{!}{\includegraphics{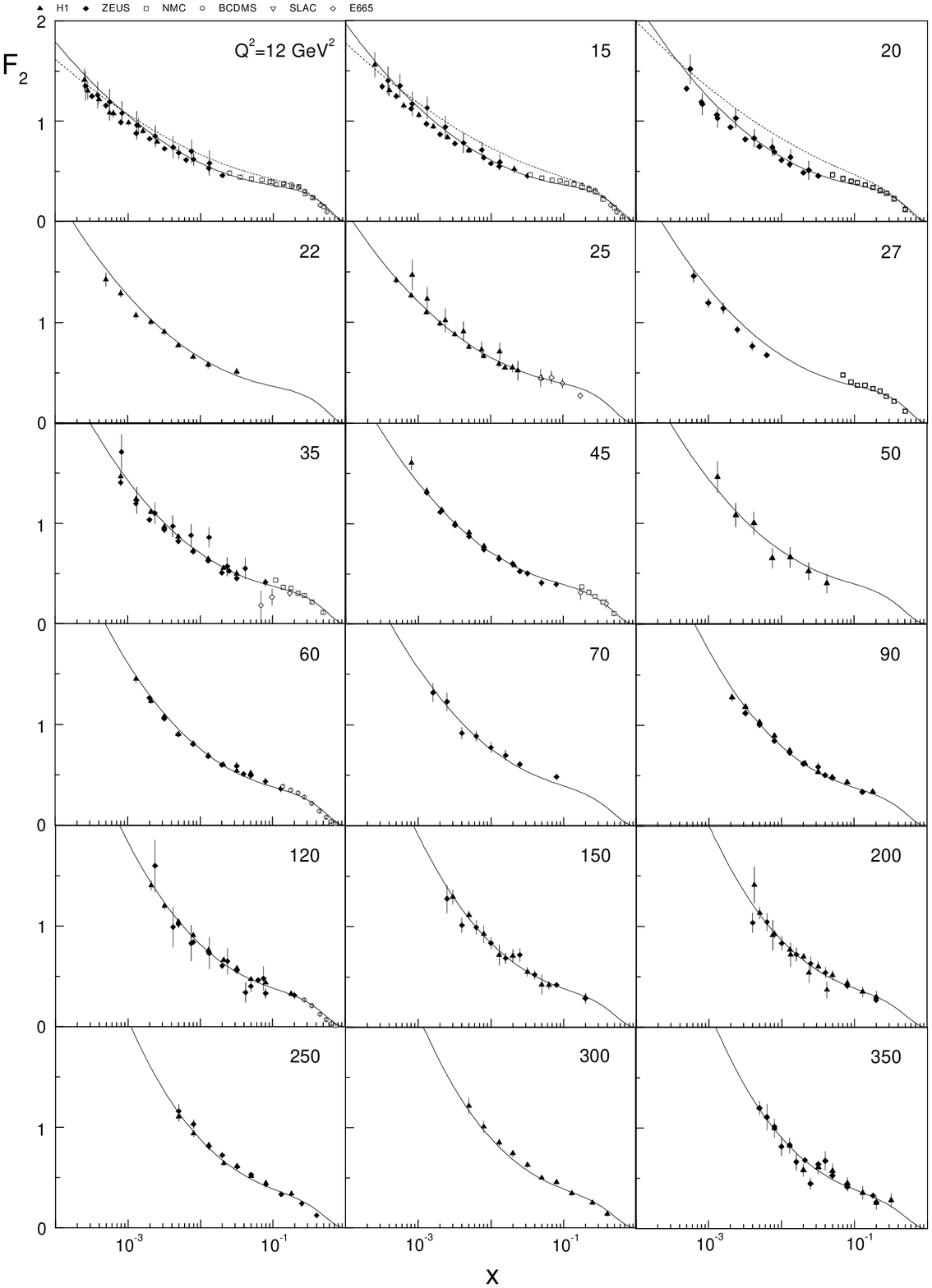}}
\caption{The same as in Fig. 2.} \label{fig:3}
\end{figure*}

\begin{figure*}
\resizebox{0.9\textwidth}{!}{\includegraphics{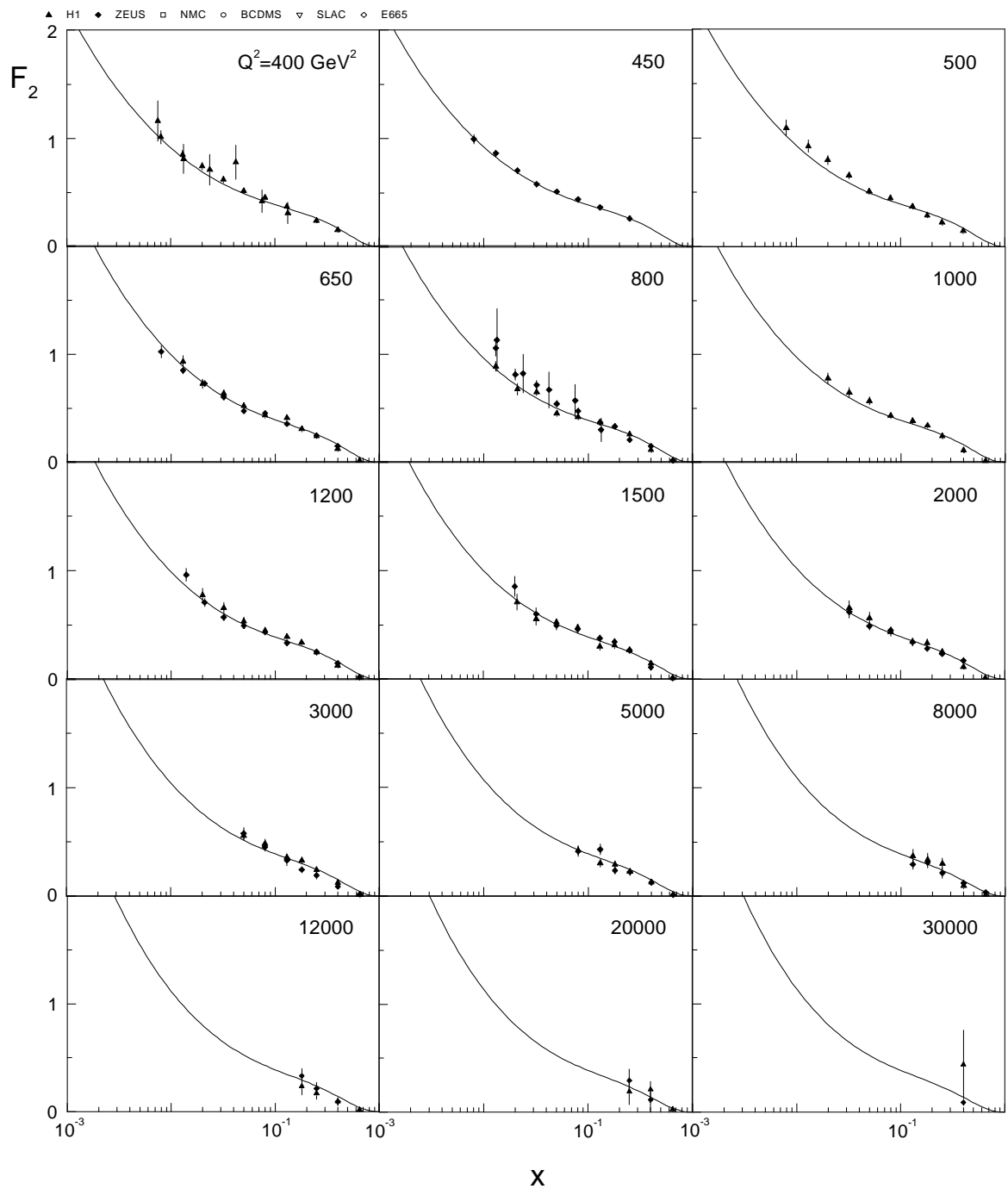}}
\caption{Results for the evolution of $F_{2}$.} \label{fig:4}
\end{figure*}

\end{document}